\documentclass[a4paper]{jpconf}

\usepackage{graphicx,cite}

\usepackage{epsfig,multicol,multirow,cite}
\usepackage{amsmath,subfigure,latexsym,amssymb}

\newcommand{\be}{\begin{equation}}
\newcommand{\ee}{\end{equation}}

\newcommand{\bea}{\begin{eqnarray}}
\newcommand{\eea}{\end{eqnarray}} 

\newcommand{\la}{\langle}
\newcommand{\ra}{\rangle}

\newcommand{\Z}{\mathbb{Z}}
\newcommand{\R}{{\kern+.25em\sf{R}\kern-.78em\sf{I} \kern+.78em\kern-.25em}}
\newcommand{\RR}{{\kern+.25em\sf{R}\kern-.6em\sf{I} \kern+.6em\kern-.25em}}
\newcommand{\N}{{\kern+.25em\sf{N}\kern-.78em\sf{I} \kern+.78em\kern-.25em}}
\newcommand{\C}{{\kern+.25em\sf{C}\kern-.50em\sf{I} \kern+.50em\kern-.25em}}

\newcommand{\shi}{\hspace{1.95mm}}

\makeatletter
\@addtoreset{equation}{section}
\makeatother

\begin{document}
\title{Topology in the 2d Heisenberg Model \newline under Gradient Flow}

\author{I O Sandoval$^{1}$, W Bietenholz$^{1}$, P de Forcrand$^{2,3}$, 
  U Gerber$^{1,4}$ \newline and H Mej\'{\i}a-D\'{\i}az$^{1}$}

\address{$^{1}$  Instituto de Ciencias Nucleares,
Universidad Nacional Aut\'{o}noma de M\'{e}xico, A.P.\ 70-543,\\
\hspace*{1.5mm} C.P.\ 04510 Ciudad de M\'{e}xico, Mexico}

\address{$^{2}$ Institut f\"{u}r Theoretische Physik, ETH Z\"{u}rich,
  Wolfgang-Pauli-Str.\ 27, CH--8093 Z\"{u}rich, \\
\hspace*{1.5mm}  Switzerland}

\address{$^{3}$ CERN, Theory Division,
  CH-1211 Gen\`{e}ve 23, Switzerland}

\address{$^{4}$ Instituto de F\'{\i}sica y Matem\'{a}ticas,
  Universidad Michoacana de San Nicol\'{a}s de Hidalgo, \\
\hspace*{1.5mm}  Edificio C-3,
  Apdo.\ Postal 2-82, C.P.\ 58040, Morelia, Michoac\'{a}n, Mexico}

\ead{ilya.orson@ciencias.unam.mx}

\begin{abstract}
  The 2d Heisenberg model --- or 2d O(3) model --- is popular
  in condensed matter physics, and in particle physics as a toy model
  for QCD. Along with other analogies, it shares with 4d Yang-Mills
  theories, and with QCD, the property that the configurations are
  divided in topological sectors. In the lattice regularisation
  the topological charge $Q$ can still be defined such that
  $Q \in \Z$. It has generally been observed, however, that the
  topological susceptibility $\chi_{\rm t} = \la Q^2 \ra /V$
  does not scale properly in the continuum limit, {\it i.e.}\ that
  the quantity $\chi_{\rm t} \xi^{2}$ diverges for $\xi \to \infty$
  (where $\xi$ is the correlation length in lattice units). Here
  we address the question whether or not this divergence persists
  after the application of the Gradient Flow.
\end{abstract}

\section{The 2d O(3) model on the lattice}

We consider square lattices of volume $V= L \times L$, and we refer
to lattice units, {\it i.e.}\ the spacing between lattice
sites is set to $1$. At each site $x$ there is a 3-component
classical spin variable of length 1, $\vec e_{x} \in S^{2}$.
The standard lattice action of a configuration $[\vec e \, ]$
is given by
\be
S [\vec e \, ] = \beta \sum_{\la xy\ra }
(1 - \vec e_{x} \cdot \vec e_{y}) \ , 
\ee
where the sum runs over the nearest neighbour lattice sites.
We assume periodic boundary conditions and $\beta >0$.
Obviously, this model is symmetric under global O(3) spin rotations.

In solid state physics this represents a model for a ferromagnet.
Its r\^{o}le as a toy model for QCD is based on asymptotic
freedom \cite{Poly}, a dynamically generated mass gap (which was
computed with the Bethe ansatz \cite{massgap}), and the
existence of topological sectors.

\section{Monte Carlo simulation}

Since the action is real positive for any configuration,
$S [\vec e \, ] \geq 0$, it can be employed to define a probability
\be
p[\vec e \, ] = \frac{1}{Z} \, e^{-S [\vec e \, ]} \ , \quad
  Z = \int D \vec e \ e^{-S [\vec e \, ]} \ .
\ee
It is normalised by the partition function $Z$, which is
given by a functional integral over all configurations.

A Monte Carlo simulation generates a large set
of random configurations with this probability distribution, which
enable numerical measurements. To this end, we used the highly efficient
cluster algorithm \cite{Wolff89}, both in its single-cluster and its
multi-cluster version. It is far superior to local update algorithms,
which suffer {\it e.g.}\ from a very long auto-correlation time with
respect to the topological charge $Q$, in particular close to criticality.

\section{Scale and parameters}

As usual, the intrinsic scale of the system is given by its
correlation length $\xi$. It describes the decay of the
correlation function, which can be computed as 
the correlation between layer averages,
\be  \label{corfun}
\la \vec s_{x_2} \cdot \vec s_{y_2} \ra \propto
\cosh \left( - \frac{|x_2 - y_2 | - L/2}{\xi} \right) \ ,
\quad \vec s_{x_2} = \frac{1}{L} \sum_{x_1} \vec e_{x} \ ,
\quad x= (x_{1},x_{2}) \ .
\ee
This proportionality relation holds if the size $L$ is large
compared to $\xi$, and the numerator $|x_2 - y_2 | - L/2$
is sufficiently small. We determined $\xi$ by a fit
in the interval $L/3 \leq |x_2 - y_2 | \leq 2L/3$, as suggested
in Ref.\ \cite{Wolff90}. For a variety of parameters, our results
for $\xi$ are consistent with values given in the literature,
for instance in Refs.\ \cite{Wolff90,ABF,Kim}.

The correlation length depends essentially on the parameter $\beta$,
and to some extent also on the size $L$. As we vary $L$ at fixed
$\beta$, $\xi$ is asymptotically stable in large volumes, $L \gg \xi$.
Generally, the {\em finite-size effects} are suppressed by the ratio
$L / \xi$. Our study was performed in boxes of constant size,
\be  \label{Lxi}
L \simeq 6 \, \xi \ ,
\ee
which suppresses the finite-size effects quite well.
This required a fine-tuning of $\beta$ in each volume.

On the other hand, the {\em lattice artifacts} depend on the
ratio of the correlation length and the lattice spacing.
Since we are using lattice units, this ratio is simply given
by $\xi$. Our values of $\beta$ and $\xi$, in the range
$L = 24 \dots 404$, are listed in Table \ref{tabdata},
in the appendix.

Hence we study the convergence to the continuum limit by increasing
$L$, and increasing $\beta$ accordingly such that $L/\xi \simeq 6$
persists. This amounts to an amplification of $\xi$ at small
finite-size effects, which are kept of the same magnitude, {\it i.e.}\
we perform a controlled extrapolation towards the continuum.

\section{Topological charge and susceptibility}

\begin{figure}[h!]
  \begin{center}
    \includegraphics[width=9pc]{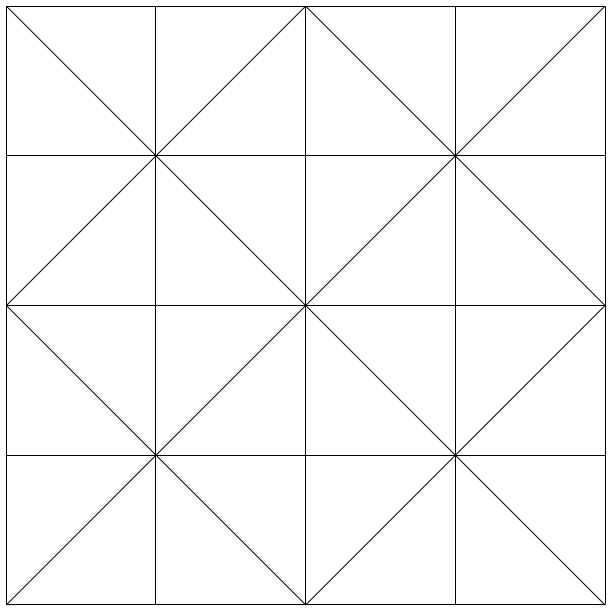}
    \hspace*{7mm}
    \includegraphics[width=13pc]{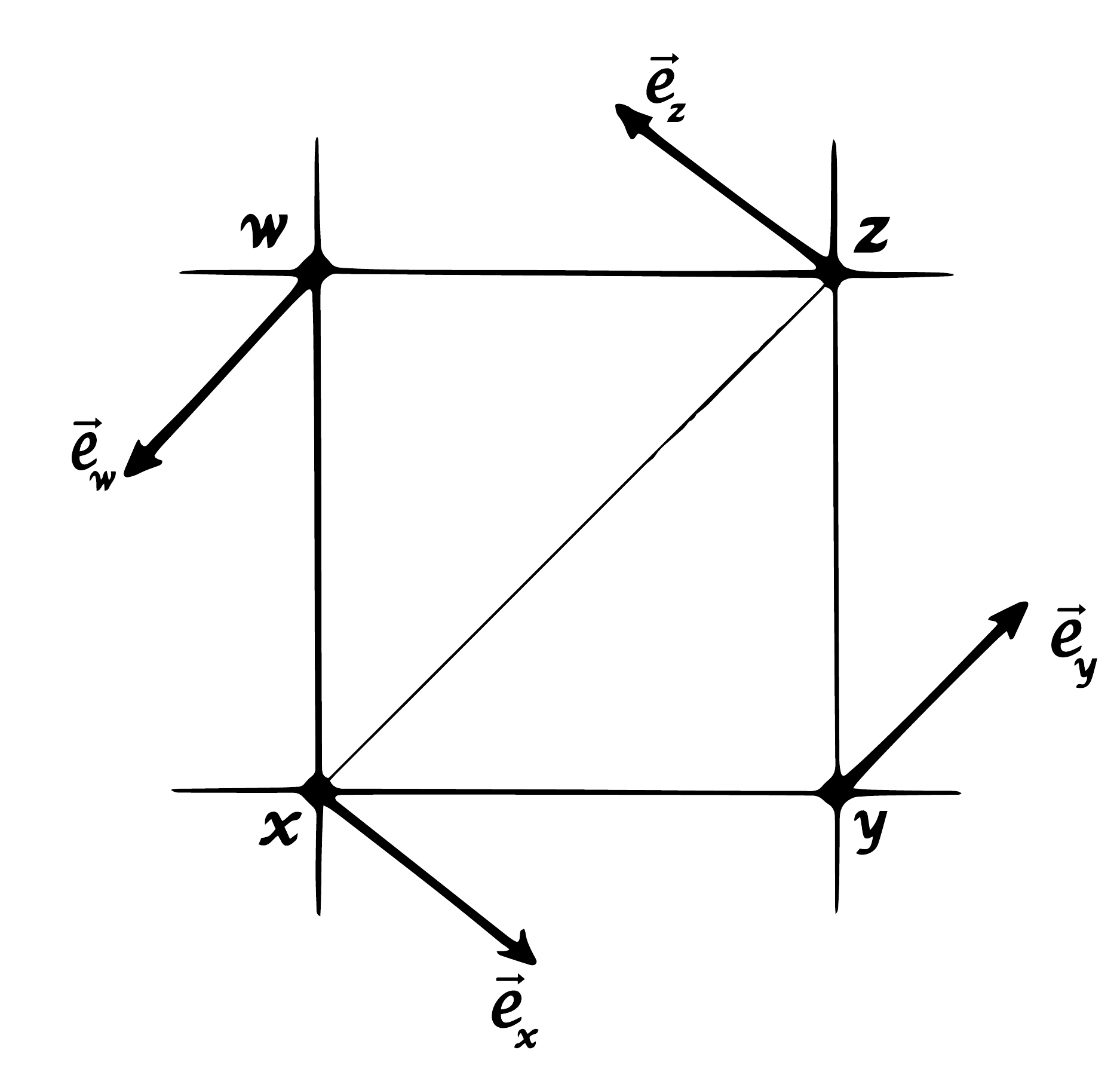}
  \end{center}
  \caption{On the left: illustration of the division of the lattice
    plaquettes into triangles, in an alternating order.
    On the right: the spins at the vertices of each triangle ---
    here $(\vec e_{x}, \vec e_{y}, \vec e_{z})$ and
    $(\vec e_{x}, \vec e_{z}, \vec e_{w})$ --- span a spherical
    triangle, with the oriented minimal areas $A_{xyz}$ and
    $A_{xzw}$, respectively. Their sum defines the topological charge
    density at the plaquette under consideration.}
\label{Qgeofig}
\end{figure}

The geometric definition of the topological charge $Q$
\cite{BergLuscher} of a lattice configuration $[\vec e \, ]$
has the virtue that it provides integer values,
$Q [\vec e \, ] \in \Z$ for all configurations (up to a
subset of measure zero). We split each plaquette into two triangles,
in an alternating order, as illustrated in Figure \ref{Qgeofig}
on the left.
The spins at the vertices of one triangle, say $(\vec e_{x},
\vec e_{y}, \vec e_{z})$, span a spherical triangle on $S^{2}$.
We refer to the spherical triangle with minimal area, and a
fixed orientation (which determines the sign). This oriented
area $A_{xyz}$ defines the topological charge density
$A_{xyz}/4 \pi$, and therefore the winding number, or
topological charge
\be
Q [\vec e \, ] = \frac{1}{4 \pi} \sum_{\la xyz \ra} A_{xyz} \in \Z \ ,
\ee
where the sum runs over all triangles (for obtaining integer $Q$-values,
it is crucial to account for the periodic boundary conditions). The
explicit formulae are given in Refs.\ \cite{BergLuscher,topact,chitfix}.

\begin{figure}[h!]
  \begin{center}
    \includegraphics[width=25pc,angle=0]{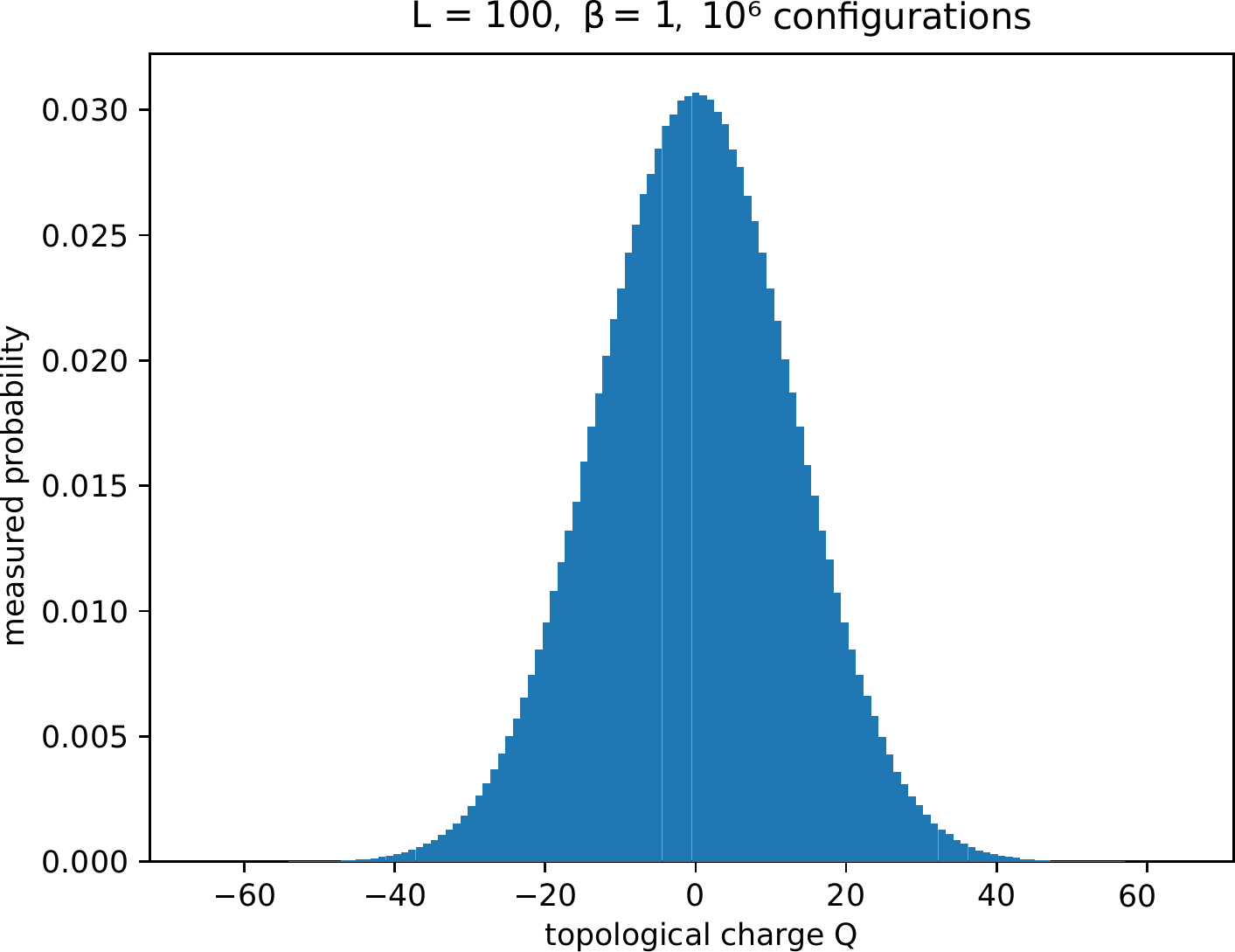}    
  \end{center}
  \caption{Histogram of the topological charges, obtained at $L=100$,
    $\beta =1$ with a statistics of $10^{6}$ configurations. Its shape is
  approximately Gaussian, with a width of $ \la Q^{2}\ra  =\chi_{\rm t}V$.}
\label{Qhisto}
\end{figure}

Parity symmetry implies $\la Q \ra =0$, hence the topological
susceptibility takes the form
\be
\chi_{\rm t} = \frac{1}{V} \left( \la Q^{2} \ra - \la Q \ra^{2} \right)
= \frac{ \la Q^{2} \ra}{V} \ .
\ee
Figure \ref{Qhisto} shows an example for a histogram of the
topological charge distribution, which tends to be approximately
Gaussian.\footnote{Results for the kurtosis term
$c_{4} = \left( \la Q^{2}\ra^{2} - \la Q^{4} \ra \right)/V$,
  as a measure for the deviation from a Gaussian distribution,
  are given in Ref.\ \cite{slabLat16}.}

For the topological susceptibility to have a sound continuum
limit, in a large volume, the (dimensionless) physical quantity
$\chi_{\rm t} \xi^{2}$ should converge to a finite constant,
\be
\lim_{\xi \to \infty} \ \chi_{\rm t} \xi^{2} = {\rm constant} \ .
\ee
The question whether this is actually the case has been debated
since the 1980s. While it was controversial for a while ---
based on considerations of various lattice actions and definitions
of the lattice topological charge ---
the consensus is now that this limit diverges, {\it i.e.}\ the
topology of this model is not well-defined in the continuum limit.
This appears as a conceptual disease of the 2d O(3) model.

After the first numerical evidence for this divergence \cite{BergLuscher},
a semi-classical argument was elaborated in Ref.\ \cite{Luscher82}: it
considers very small topological windings of a lattice configuration
(``dislocations''). For increasing $\beta$ they are suppressed, but
the semi-classical picture suggests that this suppression is not
sufficient to compensate for the entropy growth due to the
increase in $\xi$.

Later a sophisticated version of a (truncated) classically perfect
lattice action was applied, which suppresses such dislocations
by numerous additional couplings, beyond nearest neighbour sites
\cite{Blatter}. However, the numerical results with this action
(which were also obtained at $L \simeq 6 \, \xi$) suggest that the
term $\chi_{\rm t} \xi^{2}$ still does not converge to a finite value
in the continuum limit. That study observed a logarithmic divergence
of $\chi_{\rm t} \xi^{2}$ with $\xi$.

\section{Gradient Flow}

In recent years, the Gradient Flow has attracted considerable
attention in the lattice community. This interest was boosted in
particular by Refs.\ \cite{LuscherGF}. Unlike previously popular
methods, where lattice configurations were smoothened {\it ad hoc}, 
the Gradient Flow performs such a smoothing in a controlled manner,
which corresponds to a renormalisation group flow. When applied
to the 2d O(3) model, one could intuitively
imagine that it removes the (small) dislocations, while preserving
topological winding on a large scale --- in the semi-classical
simplification they are represented by instantons with large radii.
Hence, the question arises if the application of the Gradient
Flow leads to a finite continuum limit of $\chi_{\rm t} \xi^{2}$.

The formula for the Gradient Flow in the 2d O($N$) models
has been written down in Ref.\ \cite{MaSu}. Here we reproduce
it for the reader's convenience. In the continuum, the spin
components $e(x)^{i}, \ i=1 \dots N,$ are modified according
to the differential equation
\be
\partial_{t} \, e(t,x)^{i} = P^{ij}(t,x) \ \Delta \, e(t,x)^{j} \ , \quad
P^{ij}(t,x) = \delta^{ij} - e(t,x)^{i} \, e(t,x)^{j} \ ,
\ee
where $t$ is the Gradient Flow time (which generically has the
dimension [length]$^{2}$) starting at $t=0$, and $\Delta$ is the
Laplace operator. On the lattice we replace $\vec e(t,x)$
by the spin variable at one site, $\vec e(t)_{x}$, and we
apply the standard discretisation of the Laplacian,
\be
\Delta e(t,x)^{j} \longrightarrow
e(t)_{x_{1}+1,x_{2}}^{j} + e(t)_{x_{1},x_{2}+1}^{j} +
e(t)_{x_{1}-1,x_{2}}^{j} + e(t)_{x_{1},x_{2}-1}^{j} - 4 e(t)_{x_{1},x_{2}}^{j} \ .
\ee
For the corresponding spin rotations we apply the Runge-Kutta
4-point method.
In practice we proceed as follows: for a given configuration,
the gradients are computed for all spin variables, at the
flow time instants which are needed for the Runge-Kutta scheme.
This is done in a fixed configuration; then all the spins
are simultaneously modified with a Gradient Flow time step
of $dt = 10^{-4}$. (After each step, the normalisation of
the modified spins is re-adjusted.)

This value of $dt$ seems to be sufficiently small to avoid
significant artifacts due to the flow time discretisation,
whereas some discretisation effects were observed at
$dt = 10^{-3}$. On the other hand, for $dt = 10^{-4}$ we did not
find any significant difference when we modified the spins one by
one lexicographically.

In order to explore the effect of the Gradient Flow on the topology
towards the continuum limit, we have to set a scale for the flow time.
Thus the results at various $L$ and $\beta$ can be related.
We follow the recipe of Refs.\ \cite{LuscherGF} by considering
the energy density. In our case, it is calculated as
\be
E_{x} = 4 - \vec e_{x} \cdot ( \vec e_{x_{1}+1,x_{2}} + \vec e_{x_{1},x_{2}+1} 
+ \vec e_{x_{1}-1,x_{2}} + \vec e_{x_{1},x_{2}-1} )
\ee
at some lattice site $x$ (it vanishes for a uniform configuration,
and grows the more the spin directions differ). Its expectation
value is trivially related to the mean value of the action density,
$\la E \ra = \la S \ra /(\beta V)$. In QCD, L\"{u}scher suggested
to choose the Gradient Flow time unit $t_{0}$ such that
$\la E \ra \, t_{0}^{2} = 0.3$ \cite{LuscherGF}. In a 2d theory the
corresponding dimensionless product reads $\la E \ra \, t$, and we
had to fix a lower reference value, which we chose as
\be  \label{t0eq}
\la E \ra \, t_{0} = 0.08 \ .
\ee
As the flow time proceeds, the term $\la E \ra \, t$ rises from 0
to some maximum before gradually decreasing again. If the reference
value is taken too large, it is not even attained for all parameter
sets in our study (if one increases $L$ and $\beta$ more and more,
this maximum decreases monotonously).
The above reference value of $0.08$ captures lattice sizes up
to $L=606$ \cite{prep}, where $\beta$ is always tuned such that
relation (\ref{Lxi}) holds.
\begin{figure}[h!]
  \vspace*{-4mm}
  \begin{center}
    \includegraphics[width=28pc,angle=0]{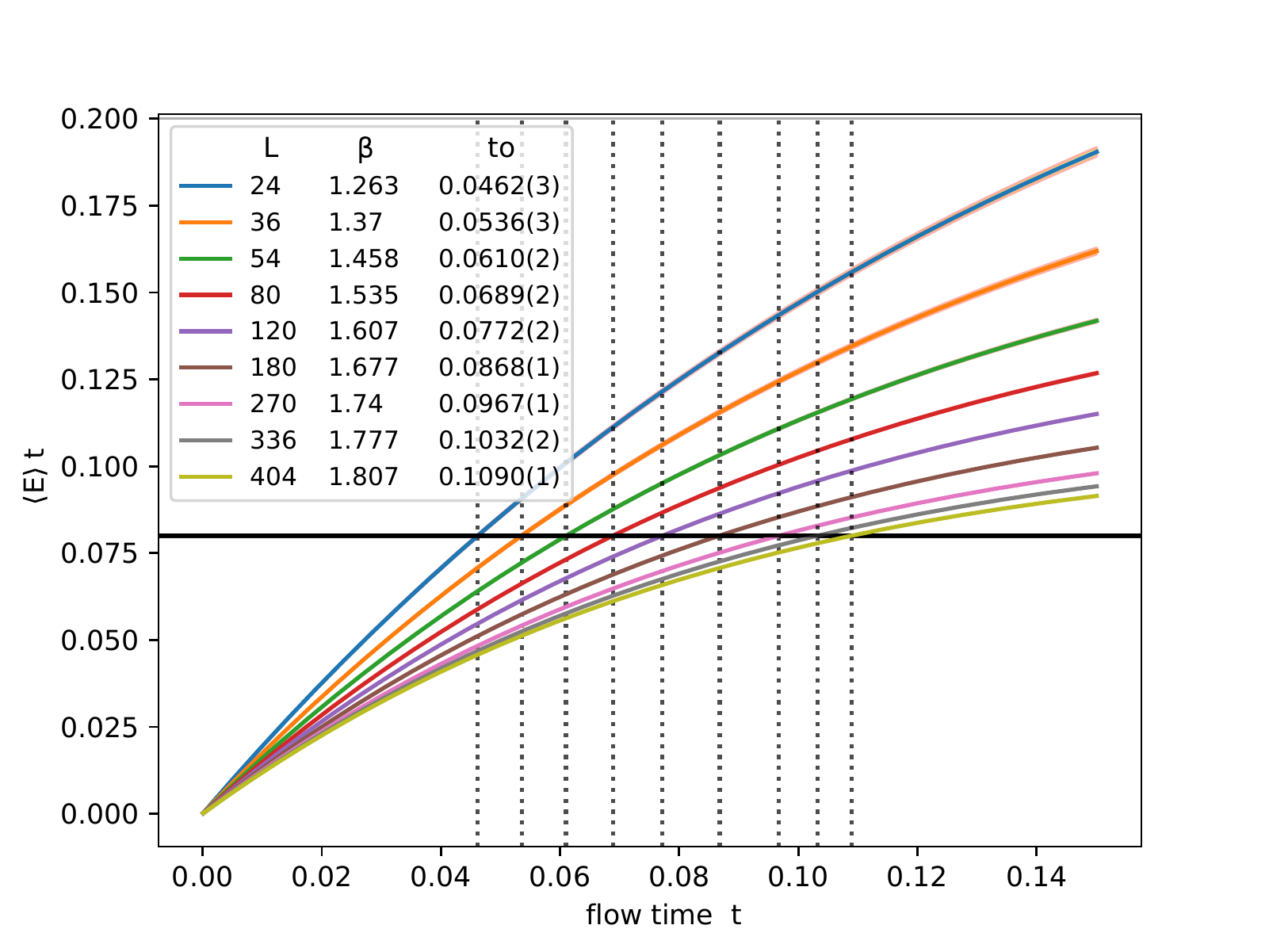}
  \vspace*{-2mm}
  \end{center}
  \caption{The Gradient Flow time evolution of the dimensionless
    term $\la E \ra \, t$, for a variety of lattice sizes
    $L = 24 \dots 404$, always with $\xi \simeq L/6$.
    We illustrate in particular the
    time $t_{0}$, where the reference value of $0.08$ is attained
    for the first time, so it matches the condition (\ref{t0eq}).}
\label{GFtimeunit}
\end{figure}
Figure \ref{GFtimeunit} shows the evolution of the term
$\la E \ra \, t$ under Gradient Flow for a variety of volumes
(with the suitable $\beta$-value), and the time $t_{0}$ where
it amounts to $0.08$ for the first time (in the long-time evolution
it decreases again down to this value and below). The resulting
$t_{0}$-values are given in the plot of Figure \ref{GFtimeunit}
and in Table \ref{tabdata}; they are rather small compared to
typical values in QCD.

\begin{figure}[h!]
  \begin{center}
    \includegraphics[width=28pc,angle=0]{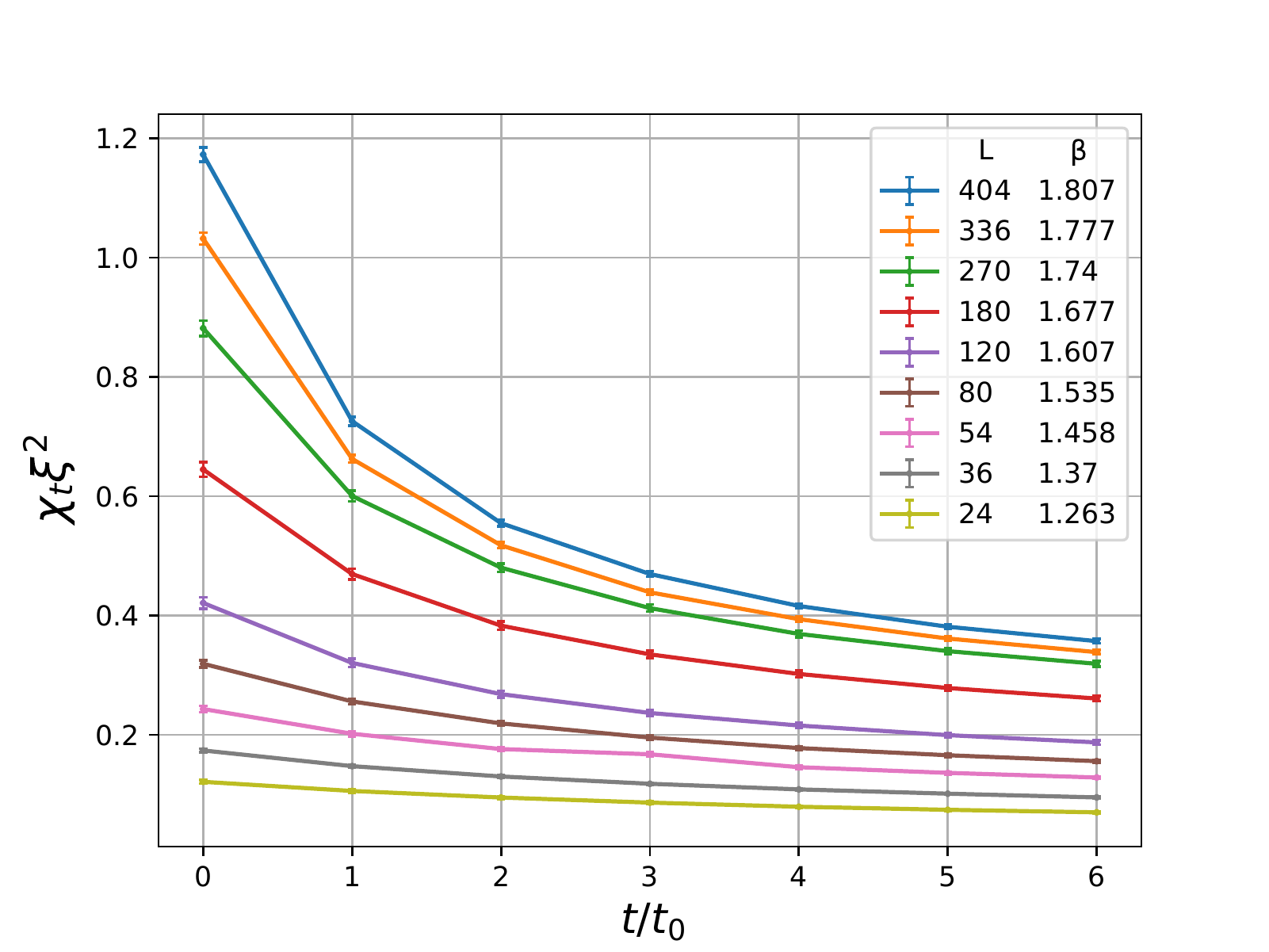}    
  \vspace*{-3mm}
  \end{center}
  \caption{The evolution of the ``scaling quantity''
    $\chi_{\rm t} \, \xi^{2}$ under the Gradient Flow, for
    several volumes, up to $6 t_{0}$. We see that this quantity
    is strongly reduced by the smoothing procedure. Since $\xi$
    remains almost constant, this demonstrates the destruction
    of a significant part of the topological windings. In this
    respect, the magnitude of our $t_{0}$-values is perfectly
    sensible.}
\label{evolfig}
\end{figure}

Next we consider the correlation function (\ref{corfun}),
\be
C(r) = \la \vec s_{x_{2}} \cdot  \vec s_{x_{2}+r} \ra \ .
\ee
It coincides with the connected correlation function due to the
O(3) rotation symmetry, which implies $\la \vec s_{x_{2}} \ra = \vec 0$.
\begin{figure}[h!]
  \vspace*{-3mm}
  \begin{center}
    \includegraphics[width=28pc,angle=0]{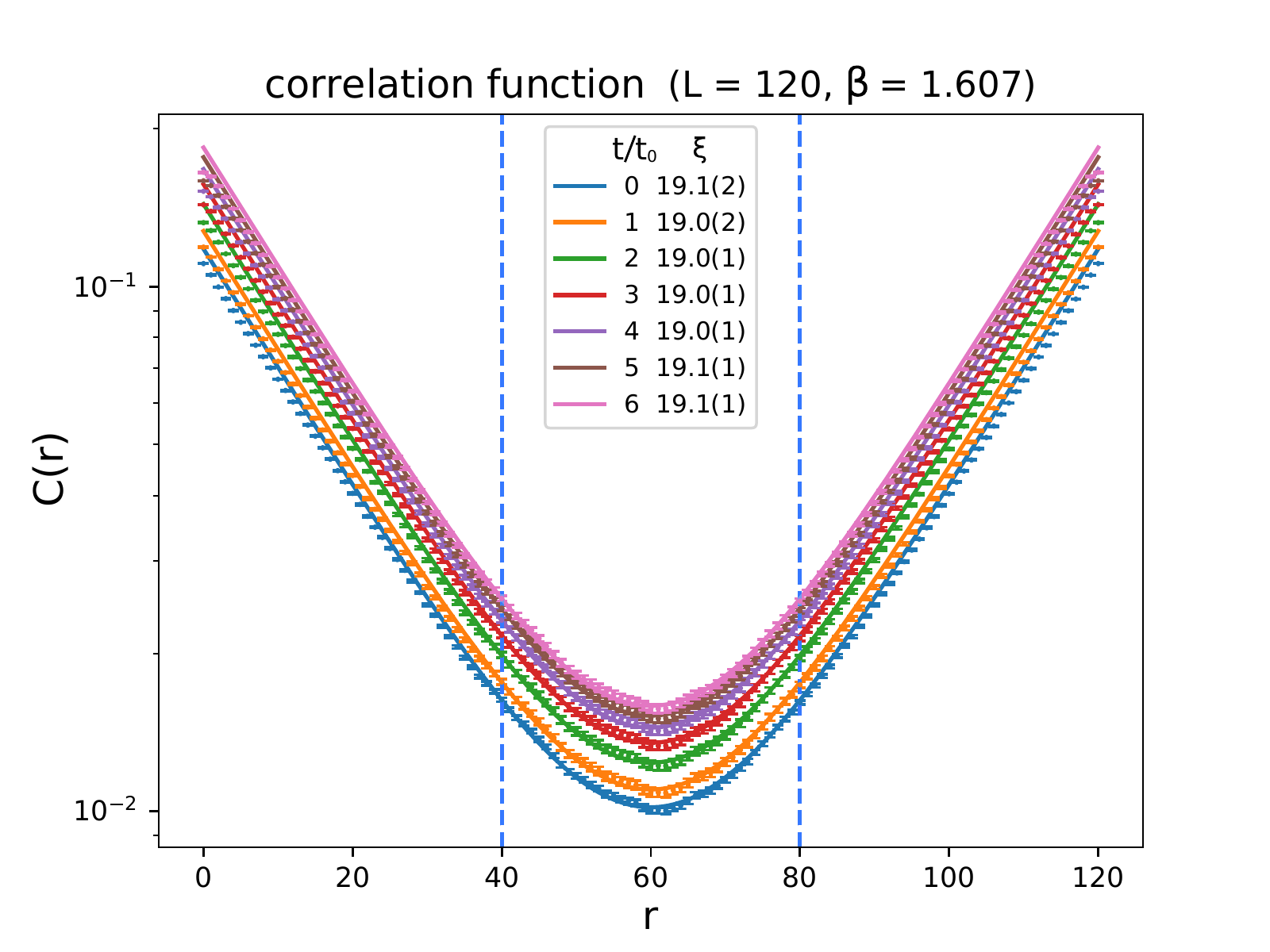}    
  \end{center}
  \vspace*{-3mm}
  \caption{The correlation function $C(r)$
    at $L=120$, $\beta =1.607$ (as an example). We see that
    the correlation at a fixed layer separation $r$
    increases due to the Gradient Flow, but
    the correlation length $\xi$ remains almost unaltered,
    up to flow times as long as $6 t_{0}$. This is the generic
    behaviour that we also observed in all other volumes
    under consideration. In this regime, the value of $\xi$
    is a long-range property, which is hardly affected by the
    Gradient Flow.}
\label{corre}
\end{figure}
Figure \ref{corre} shows an example for the behaviour of the
correlation function under Gradient Flow. As the flow time
$t$ proceeds, the correlation at a fixed distance $r$
is getting stronger, as one might expect. However, when
we perform the fit to measure the correlation length $\xi$,
according to the formula (\ref{corfun}), we see that $\xi$
hardly changes. This observation holds for all parameter
sets $(L,\beta)$ that we considered. Hence the intrinsic scale
of the system is almost constant under the Gradient Flow, at least
up to about $6 \, t_{0}$, although flow times of this magnitude
smoothen the configurations significantly over distances below
$\xi$, as we will see in Figure \ref{evolfig}.

Now we consider the quantity $\chi_{\rm t} \, \xi^{2}$, which is
supposed to be the scaling term towards the continuum limit,
as we mentioned before. Figure \ref{evolfig} shows 
that this term is suppressed significantly for the
flow times in our study --- in our three largest volumes
($L \geq 270$), $t=3t_{0}$ already reduces $\chi_{\rm t} \, \xi^{2}$
below half of its initial value.
Since $\xi$ hardly changes, this reduction is due to the
destruction of topological windings. This seems compatible
with the picture of the elimination of dislocations. The
final question is whether this effect is sufficient to entail
a finite continuum limit of $\chi_{\rm t} \, \xi^{2}$, after
a fixed multiple of the flow time unit $t_{0}$.

\section{Continuum limit}
\label{sec:contlim}

According to L\"{u}scher, in QCD any finite amount of Gradient Flow
removes the UV divergences of the original theory \cite{LuscherGF}.
This motivates us to investigate whether the same effect takes
place in the 2d O(3) model, such that the Gradient Flow cures its
topological UV behaviour.

\begin{figure}[h!]
  \begin{center}
    \includegraphics[width=30pc,angle=0]{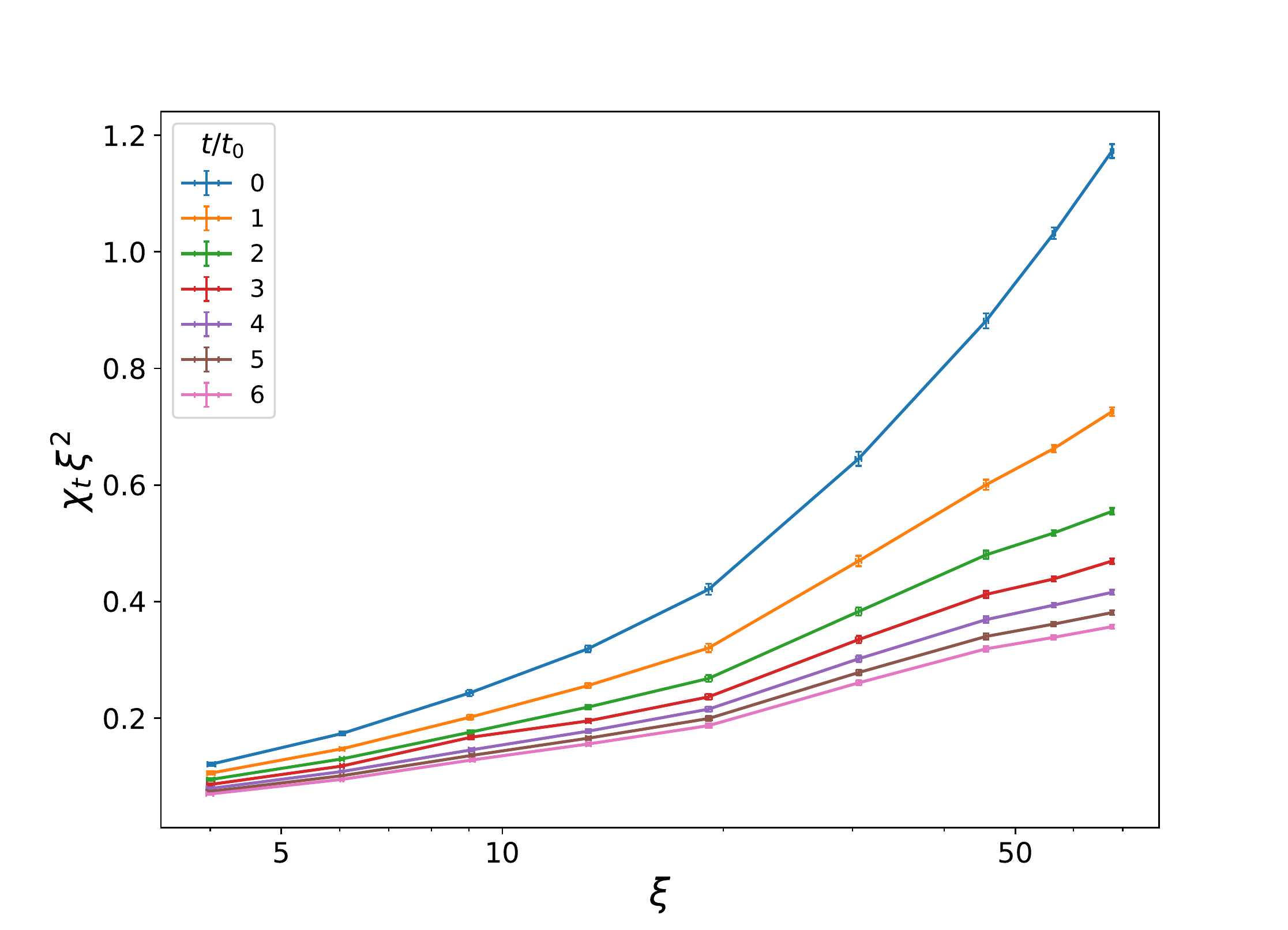}    
  \vspace*{-6mm}
  \end{center}
  \caption{The ``scaling quantity'' $\chi_{\rm t} \, \xi^{2}$, at flow
    times $t/t_{0} = 0 \dots 6$, and correlation length
    $\xi \simeq 4 \dots 67.3$ (at $L/\xi \simeq 6$).
    Thus our study advances up to very fine lattices, but
    $\chi_{\rm t} \, \xi^{2}$ still does not seem to scale towards
    a finite continuum limit at fixed $t/t_{0}$.}
\vspace*{-3mm}
\label{finale}
\end{figure}

Hence we finally arrive at the crucial question how $\chi_{\rm t} \, \xi^{2}$
behaves when we approach the continuum limit. This behaviour is
shown in Figure \ref{finale}, based on $10^{5}$ configurations
in each volume, and part of the data are given
in Table \ref{tabdata}. Our study extends up to $\xi \simeq 67.7(3)$,
which is close to the continuum limit indeed, but we cannot see any
trend towards a convergence of $\chi_{\rm t} \, \xi^{2}$ to a
finite value, at any fixed ratio $t/t_{0} = 1,2 \dots 6$.

At $t=0$ the data are very well compatible with a logarithmic
divergence of the form
$\chi_{\rm t} \, \xi^{2} = c_{1} \ln (c_{2} \xi + c_{3})$
(where $c_{i}$ are constants), as it was observed before
for a classically perfect action \cite{Blatter}, and for two types
of topological lattice actions \cite{topact}. After application of the
Gradient Flow, the quality of the fits to this function decreases somewhat.
Figure \ref{fitfig} shows the data along with the logarithmic fits
at flow time $t=0,\, 2t_{0},\, 4t_{0}$ and $6t_{0}$. 
For comparison, we considered another 3-parameter fit to a power-law
of the form $\chi_{\rm t} \, \xi^{2} = c_{1} \xi^{c_{2}} + c_{3}$.
At $t=0$ it is excellent too, but after the Gradient Flow it is
a little worse than the logarithmic fits. Table \ref{chi2dof}
displays the $\chi^{2}/{\rm d.o.f}$-values for both fitting functions.
In particular, at $t=0$ a power-law cannot be ruled out by the
present data, although a logarithmic divergence is expected.
We hope for the extension of this study to even larger volumes
\cite{prep} to be helpful also in this regard.

\section{Conclusions}

The outcome of our study is illustrated in Figure \ref{finale},
and the most important data are given in Table \ref{tabdata}.
After applying the Gradient Flow, with a fixed ratio $t/t_{0}$
(where $t_{0}$ has be determined according to eq.\ (\ref{t0eq})),
the quantity $\chi_{\rm t} \, \xi^{2}$ is reduced, which reveals
the destruction of a significant part of the topological
windings. However, as $\xi$ increases,  $\chi_{\rm t} \, \xi^{2}$
still does not show any trend of a convergence towards a finite
continuum value.
Instead our data suggest a divergence in the continuum limit,
as it was observed previously without Gradient Flow
for the standard lattice action \cite{BergLuscher},
a classically perfect action \cite{Blatter},
and for topological lattice actions \cite{topact}.

We add that prominent gauge theories with topological sectors,
in particular SU($N$) Yang-Mills theories ($N \geq 2$) and QCD, suffer
--- by default --- from the same problem: if we write the lattice
topological susceptibility as $\chi_{\rm t} = \sum_{x} \la q_{0} q_{x} \ra$
(where $q$ is the topological charge density, in a conventional
formulation), one encounters a divergence, due to the point
$x=0$.\footnote{In this notation,
  the point $x=0$ also causes the divergence of $\chi_{\rm t} \xi^{2}$
  in the 2d O(3) model \cite{topact}.} However, in those models the
problem is overcome by the application of the Gradient Flow
\cite{LuscherGF}.\footnote{For alternative solutions in
those models, we refer to Refs.\ \cite{chitQCD}.}

In contrast, at this point we conclude that the topology of the
2d O(3) model seems to be ill-defined in the continuum limit, even
after the application of the Gradient Flow. However, this study is
going to be extended to even larger $L$ and $\xi$, in order
to further check this conclusion \cite{prep}.

\ack Our interest in the issue of this article
originates from a remark by Martin L\"{u}scher.
We thank him, as well as the LOC of the XXXI Reuni\'{o}n Anual
de la Divisi\'{o}n de Part\'{\i}culas y Campos de la Sociedad
Mexicana de F\'{\i}sica, where this talk was presented by IOS.
This work was supported by DGAPA-UNAM, grant IN107915,
and by the {\it Consejo Nacional de Ciencia y Tecnolog\'{\i}a}
(CONACYT) through project CB-2013/222812. The computations were
performed on the cluster of ICN/UNAM; we thank Luciano D\'{\i}az
and Eduardo Murrieta for technical assistance.

\appendix

\section{Numerical data and quality of the fits}

\begin{figure}[h!]
\vspace*{-3.5mm}
  \begin{center}
    \includegraphics[width=26pc,angle=0]{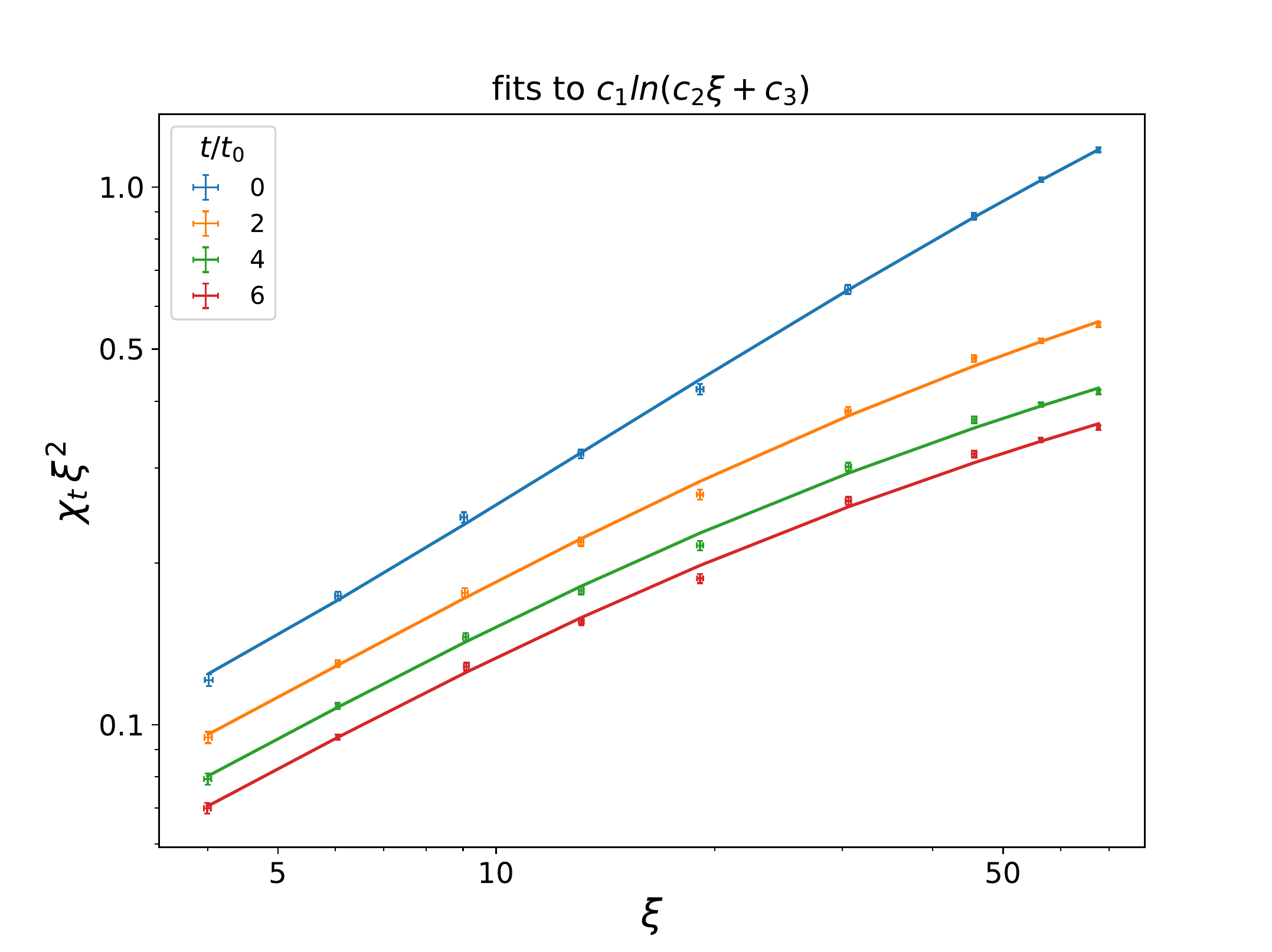}    
  \vspace*{-5mm}
  \end{center}
  \caption{The fits of our data in Figure \ref{finale}, at
    various Gradient Flow times, to the logarithmic
    ansatz $\chi_{\rm t}\, \xi^{2} = c_{1} \ln (c_{2} \xi + c_{3})$
    (where $c_{i}$ are constants). These fits work well, and they suggest
    the divergence of $\chi_{\rm t}\, \xi^{2}$ in the continuum limit.}
\label{fitfig}
\end{figure}

Table \ref{tabdata} displays the most relevant numerical results
in the lattice volumes under consideration, for the quantities
$t_{0}$, $\xi$ and $\chi_{\rm t}$.
Next we refer to our $\chi_{\rm t} \, \xi^{2}$-values as a function of
$\xi$, shown in Figure \ref{finale}. Figure \ref{fitfig} illustrates
the logarithmic fits to the data at flow time $t=0,\, 2t_{0},\, 4t_{0}$
and $6t_{0}$. Finally, Table \ref{chi2dof} gives the quality of the
fits to a logarithmic and a power-law function, as
described in the last paragraph of Section \ref{sec:contlim}.

\begin{table}[h!]
\centering
\begin{tabular}{|r|l||c||l|l||l|l|l|l|}
  \hline
$L$~ & ~~~$\beta$ & $t_{0}$ & \multicolumn{2}{c||}{$\xi$}
  & \multicolumn{4}{c|}{$\chi_{\rm t}$ \ (in units of $10^{-3})$} \\
  \hline
  & & & ~~$t=0$ & ~~$6 \, t_{0}$
  & ~~$t=0$ & ~~~$t_{0}$ & ~~$3 \, t_{0}$ & ~~$6 \, t_{0}$ \\
  \hline
  \hline
  \hline
  24  & 1.263 & 0.0462(3) & \shi4.01(5) & \shi4.00(4)
              & 7.51(4) & 6.58(4) & 5.38(3) & 4.38(2) \\
  36  & 1.37  & 0.0536(3) &  \shi6.05(5) & \shi6.04(3)
              & 4.74(3) & 4.03(2) & 3.22(2) & 2.60(1) \\
  54  & 1.458 & 0.0610(2) & \shi9.0(1) & \shi9.10(7)
              & 2.99(2) & 2.47(1) & 2.04(1) & 1.550(9)\\
  80  & 1.535 & 0.0689(2) & 13.1(1) & 13.1(1)
              & 1.86(1) & 1.495(8) & 1.139(6) & 0.907(5) \\
  120 & 1.607 & 0.0772(2) & 19.1(2) & 19.1(2)
              & 1.154(5) & 0.879(4) & 0.649(3) & 0.513(2) \\
  180 & 1.677 & 0.0868(1) & 30.6(3) & 30.6(3)
              & 0.691(3) & 0.503(2) & 0.358(2) & 0.278(1) \\
  270 & 1.74  & 0.0967(1) & 45.6(3) & 45.6(3)
              & 0.424(2) & 0.289(1) & 0.1983(9) & 0.1534(7) \\
  336 & 1.777 & 0.1032(2) & 56.4(2) & 56.3(2)
              & 0.324(1) & 0.2084(9) & 0.1381(6) & 0.1066(5) \\
  404 & 1.807 & 0.1090(1) & 67.7(3) & 67.7(3)
              & 0.256(1) & 0.1585(7) & 0.1024(5) & 0.0779(4) \\
  \hline
\end{tabular}
\caption{\label{tabdata} A summary of our numerical results in the nine
  volumes $V = L \times L$ that we investigated. In each volume, $\beta$
  was tuned such that $L / \xi \simeq 6$, and $t_{0}$ was determined
  by the condition $\la E \ra \, t_{0} = 0.08$. We see that the correlation
  length $\xi$ hardly changes under Gradient Flow, but the topological
  susceptibility is significantly reduced as we proceed up to $6 t_{0}$.
  These results are based on a statistics of $10^{5}$ configurations
  in each volume.}
\end{table}

\begin{table}[h!]
\vspace*{4mm}
\centering
\begin{tabular}{|c||c|c|c|c|c|c|c|}
\hline
  fitting function & $t=0$ & $t_{0}$ & $2t_{0}$ & $3t_{0}$ & $4t_{0}$
  & $5t_{0}$ & $6t_{0}$ \\
\hline  
$c_{1} \, \ln (c_{2} \, \xi + c_{3})$ & 1.07 & 1.34 & 1.63 &
2.20 & 1.82 & 1.89 & 1.91 \\
$c_{1} \, \xi^{c_{2}} + c_{3}$ & 1.01 & 1.58 & 1.94 & 2.32
& 2.12 & 2.16 & 2.18 \\
\hline
\end{tabular}
\caption{\label{chi2dof} The $\chi^{2}/{\rm d.o.f}$-values for two
  fits of our $\chi_{\rm t} \, \xi^{2}$-results as a function of $\xi$.
  We consider two 3-parameter fitting functions, with a logarithmic
  and a power-law divergence in the continuum limit. At Gradient
  Flow time $t=0$ both fits are excellent, but in the range
  $t = t_{0} \dots 6 t_{0}$ they become somewhat worse, in particular
  for the power-law.}
\end{table}


\section*{References}
\begin{thebibliography}{11}

\bibitem{Poly} Polyakov A M 1975
  {\it Phys.\ Lett.}\ B {\bf 59} 79

\bibitem{massgap} Hasenfratz P, Maggiore M and Niedermayer F 1990
{\it Phys.\ Lett.}\ B {\bf 245} 522

\bibitem{Wolff89} Wolff U 1989
  {\it Phys.\ Rev.\ Lett.}\ {\bf 62} 361

\bibitem{Wolff90} Wolff U 1990
{\it Nucl.\ Phys.}\ B {\bf 334} 581

\bibitem{ABF}  Apostolakis J, Baillie C F and Fox G C 1991
{\it Phys.\ Rev.}\ D {\bf 43} 2687

\bibitem{Kim} J-K Kim 1994
  {\it Phys.\ Rev.}\ D {\bf 50} 4663

\bibitem{BergLuscher} Berg B and L\"{u}scher M 1981
  {\it Nucl.\ Phys.}\ B {\bf 190} 412

\bibitem{topact} Bietenholz W, Gerber U, Pepe M and Wiese U-J
  2010 {\em JHEP} {\bf 1012} 020

\bibitem{chitfix} Bautista I {\it et al.}\ 2015
{\it Phys.\ Rev.}\ D {\bf 92} 114510

\bibitem{slabLat16} Bietenholz W, Cichy K, de Forcrand P, Dromard A
and Gerber U 2016
{\it PoS} {\bf LATTICE2016} 321

\bibitem{Luscher82}  L\"{u}scher M 1982
  {\it Nucl.\ Phys.}\ B {\bf 200} 61

\bibitem{Blatter} Blatter M, Burkhalter R, Hasenfratz P and Niedermayer F
  1996 {\it Phys.\ Rev.}\ D {\bf 53} 923

\bibitem{LuscherGF} L\"{u}scher M 2010
  {\it JHEP} {\bf 1008} 071,
  {\it PoS} {\bf LATTICE2010} 015

\bibitem{MaSu} Makino H and Suzuki H 2015	
{\it PTEP} {\bf 2015} 033B08

\bibitem{prep} Bietenholz W, de Forcrand P, Gerber U,
  Mej\'{\i}a-D\'{\i}az H and Sandoval I O, in preparation

\bibitem{chitQCD} Giusti L, Rossi G C and Testa M	2004
{\it Phys.\ Lett.}\ B {\bf 587} 157

L\"{u}scher M 2004
{\it Phys.\ Lett.}\ B {\bf 593} 296

Giusti L and L\"{u}scher M 2009
{\it JHEP} {\bf 0903} 013

L\"{u}scher M and Palombi F 2010
{\it JHEP} {\bf 1009} 110
  
\end{thebibliography}
\end{document}